\documentclass[
 aps, pra,
 amsmath,amssymb,
 12pt,
 final,
tightenlines,
 nofootinbib,
 superscriptaddress,
 ]
{revtex4}

\usepackage[utf8x]{inputenc}
\usepackage[english]{babel}
\usepackage{graphicx}
\usepackage{longtable}
\usepackage{appendix}

\def\squareforqed{\hbox{\rlap{$\sqcap$}$\sqcup$}}

\def\sq{\ifmmode\squareforqed\else{\unskip\nobreak\hfil
\penalty50\hskip1em\null\nobreak\hfil\squareforqed
\parfillskip=0pt\finalhyphendemerits=0\endgraf}\fi}

\def\degr{\hbox{$^\circ$}}

\def\utw{\smash{\rlap{\lower5pt\hbox{$\sim$}}}}

\def\udtw{\smash{\rlap{\lower6pt\hbox{$\approx$}}}}

\def\diameter{{\ifmmode\mathchoice
{\ooalign{\hfil\hbox{$\displaystyle/$}\hfil\crcr
{\hbox{$\displaystyle\mathchar"20D$}}}}
{\ooalign{\hfil\hbox{$\textstyle/$}\hfil\crcr
{\hbox{$\textstyle\mathchar"20D$}}}}
{\ooalign{\hfil\hbox{$\scriptstyle/$}\hfil\crcr
{\hbox{$\scriptstyle\mathchar"20D$}}}}
{\ooalign{\hfil\hbox{$\scriptscriptstyle/$}\hfil\crcr
{\hbox{$\scriptscriptstyle\mathchar"20D$}}}}
\else{\ooalign{\hfil/\hfil\crcr\mathhexbox20D}}%
\fi}}

\input{maik.rty}

\begin{document}

\title{Detection of pulsations and a non-stable wind in IRAS\,01005+7910, an object with the B[e]-phenomenon}
\author{\firstname{A.S.}~\surname{Miroshnichenko}}
\affiliation{University of North Carolina, Greensboro, NC~27402, USA}
\affiliation{Fesenkov Astrophysical Institute,  Almaty 050020, Kazakhstan}
\email{a_mirosh@uncg.edu}
\author{\firstname{V.G.}~\surname{Klochkova}}
\affiliation{Special Astrophysical Observatory, Nizhnii Arkhyz, 369167 Russia}
\author{V.E.~Panchuk}
\affiliation{Special Astrophysical Observatory, Nizhnii Arkhyz, 369167 Russia}
\author{E.S.~Islentieva}
\affiliation{Special Astrophysical Observatory,  Nizhnii Arkhyz, 369167 Russia}

\begin{abstract}
Comparison of the spectra obtained at different observing  dates shows significant variations in
the complex   H$\alpha$ and H$\beta$  line profiles, such as a systematic increase in their emission
peak strengths in 2025$\div$2026 compared to those observed in 2013 as well as a strong variability in their
wind component. A detected variable radial velocity $V_{\odot}$ from  $-20$ to $-68$ km\,s$^{-1}$ with a
standard deviation of the mean value of K = 8.0 km\,s$^{-1}$ can be interpreted as a result of pulsations
or the presence of a companion. The stationary positions of forbidden lines  with a mean radial
velocity of  $V_{\odot}=-51.38\pm0.26$\,km\,s$^{-1}$ are taken as an improved systemic velocity.
Overall, a set of properties of IRAS\,01005+7910 points  to its status as a post-AGB star, which has
undergone the hot-bottom burning phase.
\end{abstract}

\maketitle

\section{Introduction}

Intense optical spectroscopy of objects with the B[e] phenomenon began at the end of the 20th century.
The phenomenon shows up in the spectra of stars by the presence of permitted and forbidden emission
lines of neutral metals and ions (predominantly the CNO and Fe group), which form in a structured circumstellar medium~\citep{1998A&A...340..117L}.   Objects with the B[e] phenomenon exhibit large excesses of  IR flux, while
their optical spectra contain strong hydrogen emission lines.~\citet{1998A&A...340..117L} also note a large
range of masses and luminosities of these objects as well as evolutionary stages that range from massive
supergiants to intermediate-mass stars in transition toward the Planetary Nebula phase. Many results
from studies of objects with the B[e] phenomenon obtained over a recent decades have become the basis of
numerous papers and some reviews (e.g.,~\citep{2018MNRAS.480..320M,2019Galax...7...83K,2023Galax..11...36M}).

IRAS\,01005+7910 (hereafter IRAS\,01005) is typically classified as a high-latitude post-AGB star~\citep{2000ApJ...535..275H,2013AstL...39..619A,2015MNRAS.447.1673V}, which is an intermediate-mass
(2$\div$8$ M_\odot$) star that is undergoing a transition from the asymptotic giant branch (AGB)
to the Planetary Nebula phase. As follows from Table~1 in~\citet{2022MNRAS.516L..61O}, the central
star in IRAS\,01005 is one of the hottest candidates for this type of object.  Low-amplitude brightness
variability of the object was reported in~\citet{2013AstL...39..619A}. The ASAS~SN
database~\citep{2017PASP..129j4502K}, which contains a large array of photometric data for IRAS\,01005,
also shows a similar behavior with an average visual brightness of $V=10.95\pm0.09$~mag over the last $\sim$10~years.

A study of the object's  high-resolution optical spectrum highlighted some features which call into doubt the
previous evaluation of its evolutionary state.~In particular,~\citet{2002A&A...392..143K,2014AstBu..69..439K}
detected variations in the Balmer emission lines with P\,Cyg profiles and He{\sc i} line profiles, as well as
found many weak permitted and forbidden emission lines of atoms and ions of the CNO triad and iron group.
The chemical composition of the star’s atmosphere and an initial nitrogen excess  make the star’s post-AGB
status dubious.
In order to solve the evolutionary status problem for IRAS 01005, follow-up observations, including
high-resolution spectroscopy, are needed. Spectral monitoring conducted over several years and reported
in the latter papers showed a radial velocity instability and a velocity   gradient in the stellar atmosphere.
These authors classified IRAS\,01005 as a B1.7\,{\sc i}b star that is in agreement with the fundamental
parameters determined by the model atmosphere method (T$_{\rm eff}$=21500~K and  $\log$g=3.0).
Nevertheless, the chemical abundances of the star's atmosphere and an initial nitrogen excess make the
star's post-AGB status dubious.

In this paper, we present the results of a new stage of the high resolution spectroscopy of the hot star
in the system  IRAS\,01005, in whose spectrum all the features of the B[e] phenomenon have been found.
The methods of observations and data analysis are briefly discussed in Section~\ref{sec2}. The results and
a discussion comparing them with those obtained earlier are given in Section~\ref{sec3}. Our conclusions are
presented in Section~\ref{sec4}.

\section{\'Echelle Spectroscopy at the BTA Telescope}\label{sec2}

To search for variations in spectral features of different types, we used 5 spectra taken on random dates
with the \'echelle spectrograph NES~\citep{2017ARep...61..820P} mounted in the Nasmyth focus of the
Big Telescope Alt-azimutal (BTA) of the Special Astrophysical Observatory in the Northern  Caucasus, Russia.
The stationary location of the spectrograph provides high stability of its technical parameters. The observing
dates are shown in Table\,\ref{tab1}. The spectral resolving power of NES is R=$\lambda/\Delta\lambda \ge$ 60\,000.
The signal-to-noise ratio along the spectral order varies by a factor of $\sim$1.5. The spectral range of the data
was $\Delta\lambda=391\div680$\,nm in 2013 and  $\Delta\lambda$=470$\div$780\,nm in 2025$\div$2026. An image
slicer that  rearranged the star image into 3 slices was used to reduce light loss at the NES spectrograph
entrance slit.  The spectra taken in 2025$\div$2026 were obtained without the slicer after  reconstruction of
the spectrograph.

\begin{table}
\caption{Average heliocentric radial velocities and uncertainties for three types of spectral lines measured
      in the spectra taken on different dates. The numbers of measured lines are shown in parentheses}
\begin{tabular}{llll}
\hline
Observing     &  \multicolumn{3}{c}{\boldmath{V$_{\odot}$, km\,s$^{-1}$}} \\
\cline{2-4}
Dates      & Absorptions   & Emis-Forb  & Emis-Perm   \\ [-5pt]
\hline
 29 May 2013    &$-40.43\pm0.04$\,(84) &$-51.76\pm0.09$\,(20)&$-54.13\pm0.49$\,(12) \\ [-5pt]
 21 August 2013 &$-29.93\pm0.07$\,(38) &$-51.93\pm0.12$\,(8) &$-58.17\pm0.66$\,(9) \\ [-5pt]
11 September2025&$-41.76\pm0.15$\,(25) &$-51.57\pm0.14$\,(10)&$-55.38\pm0.35$\,(13) \\ [-5pt]
 4 October 2025 &$-21.60\pm0.17$\,(27) &$-51.12\pm0.13$\,(13)&$-49.60\pm0.30$\,(12) \\ [-5pt]
 27 January 2026&$-68.09\pm0.17$\,(20) &$-50.44\pm0.12$\,(16)&$-50.68\pm0.24$\,(14) \\
\hline
\end{tabular}
\label{tab1}
\end{table}

The NES spectrograph is equipped with a large CCD detector, which has 4608~$\times$~2048~pixels,
a pixel size of  0.0135~$\times$~0.0135 mm, and a readout noise of 1.8 electrons. Cosmic ray traces were
removed by median averaging of pairs of spectra taken consecutively. The wavelength calibration was
accomplished using a Th-Ar lamp. All data reduction procedures with  1D spectra were performed with the
latest version of the DECH20t package~\citep{2022AstBu..77..519G}. Systematic measurement errors of
heliocentric radial velocities (V$_{\odot}$) using telluric lines and interstellar lines of the Na\,{\sc i}
doublet do not exceed 0.25\,km\,s$^{-1}$ on one line, while the measurement error on broad features does not
exceed 0.5\,km\,s$^{-1}$. We note that a high accuracy of the V$_{\odot}$ measurements served as an
additional criterion for  details identification.

A list of identified features in the spectrum of IRAS\,01005 was published in Table~2
of~\citet{2002A&A...392..143K}. In order to refine and expand the list, we used the results published in~\citet{2015AstBu..70...99K,2016AstBu..71...33K,2025AstBu..80..551K}, which are based on NES
spectra of similar objects with the B[e] phenomenon taken at the BTA telescope.

\section{Main Results}\label{sec3}

\begin{figure}
\hbox{
\includegraphics[angle=0,width=0.65\textwidth,height=0.40\textheight,bb=0 0 800 800,clip]{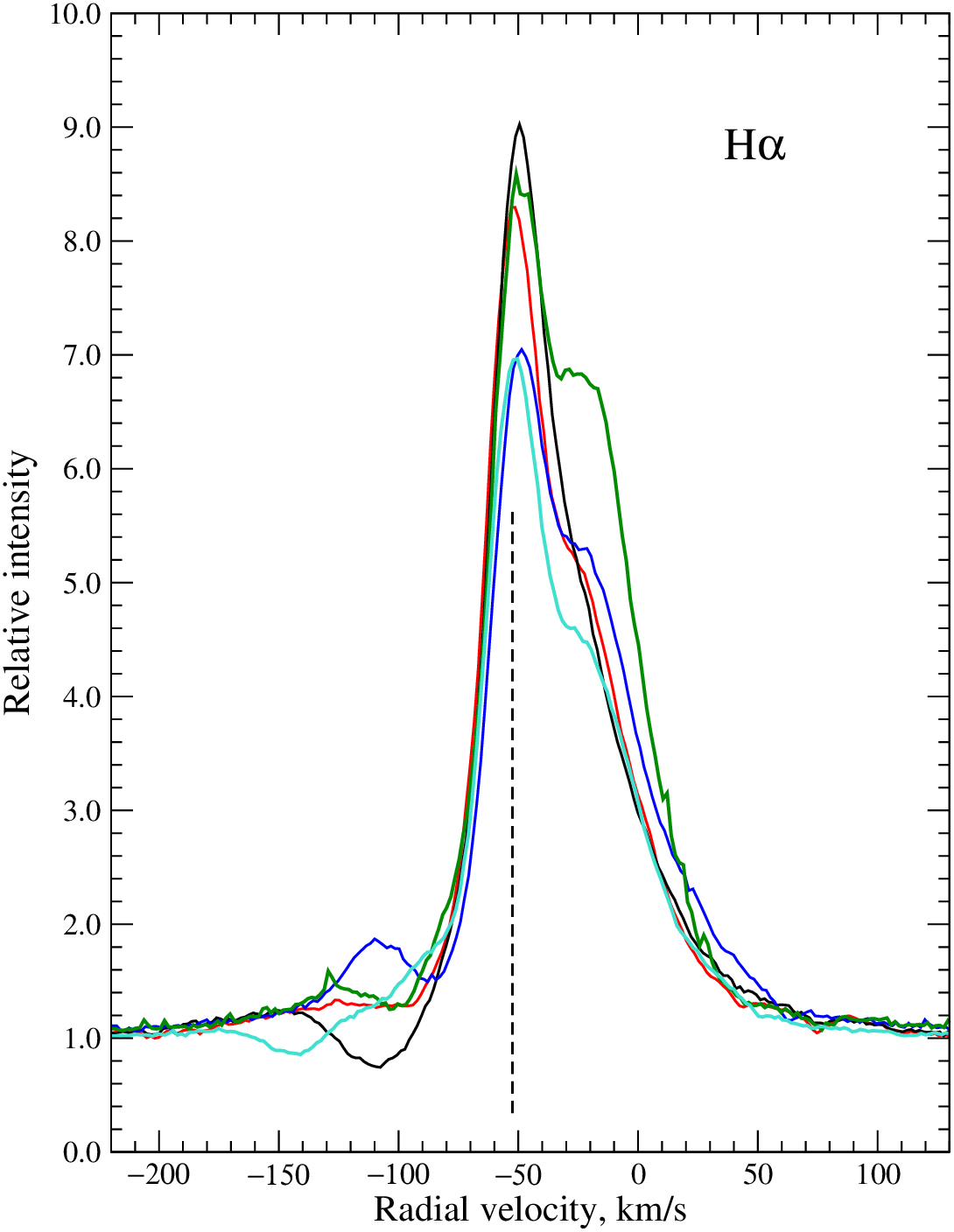}
\hspace{-2cm}
\includegraphics[angle=0,width=0.65\textwidth,height=0.46\textheight,bb=0 0 800 800,clip]{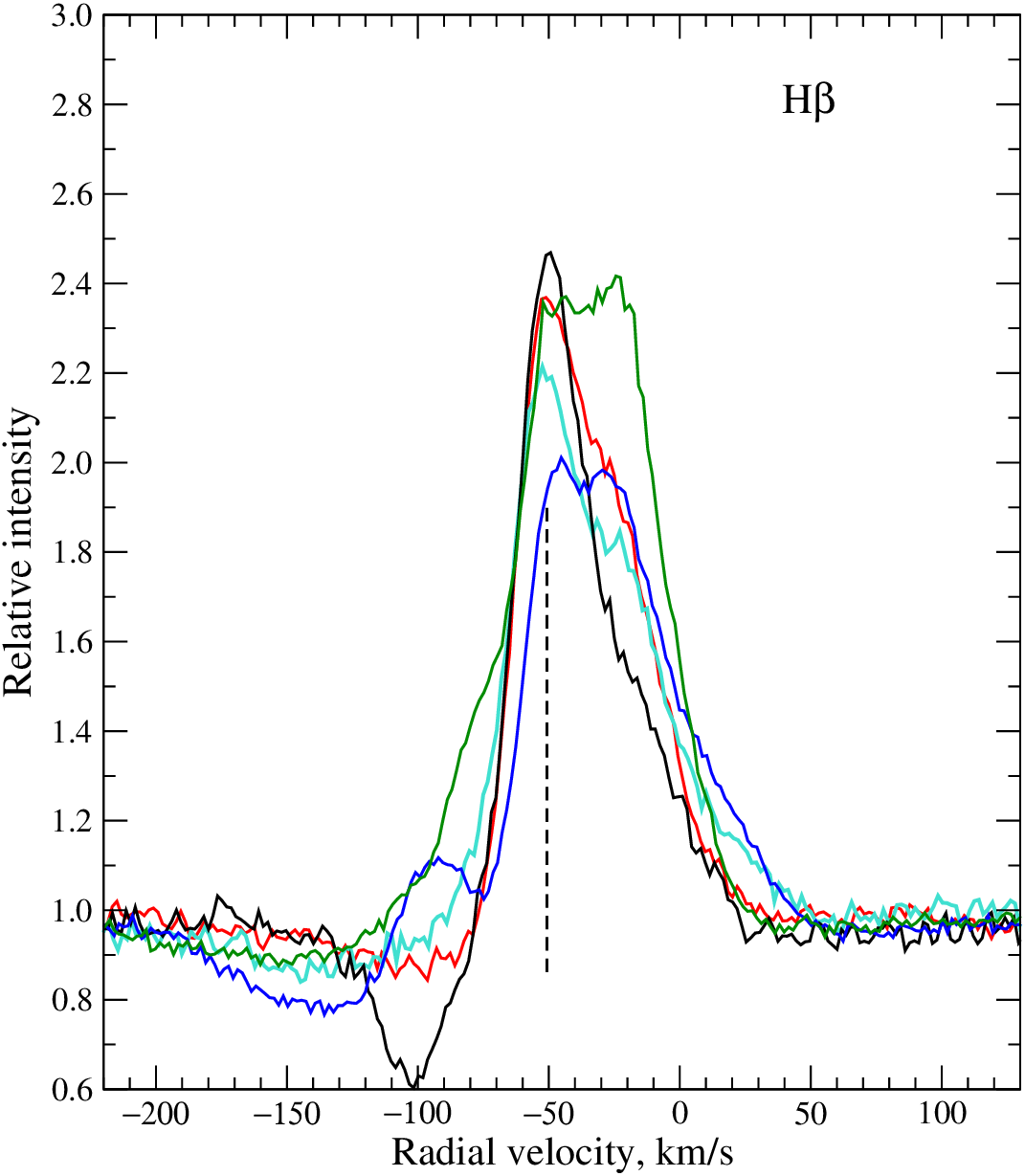}
}
\caption{H$\alpha$ and H$\beta$ line profiles taken on  29~May~2013 (blue),
21~August~2013 (green), 11~September~2025 (red),  4~October~2025 (black), and 27~January~2026 (turquoise).
The vertical dashed line corresponds to the systemic velocity V$_{\odot}$(sys)$=-51.4$ km\,s$^{-1}$.}
\label{Halpha5}
\end{figure}

The complex and variable emission lines of neutral hydrogen in the spectrum of IRAS\,01005 attract
special attention. We show examples of these line profiles in Figure~\ref{Halpha5}. The H$\alpha$ and
H$\beta$ profiles are typical for objects with the B[e] phenomenon. In particular, these P\,Cyg-type
profiles contain a strong emission component with a peak intensity up to 7--9 times the continuum level.
Figure\,\ref{Halpha5} demonstrates also the presence of a variable wind feature, suggesting a
time-dependent wind variability and/or spatial inhomogeneity of the hydrogen layer in the circumstellar
envelope.

\begin{figure}
\hbox{
\includegraphics[angle=0,width=0.45\textwidth,bb=0 0 750 600,clip]{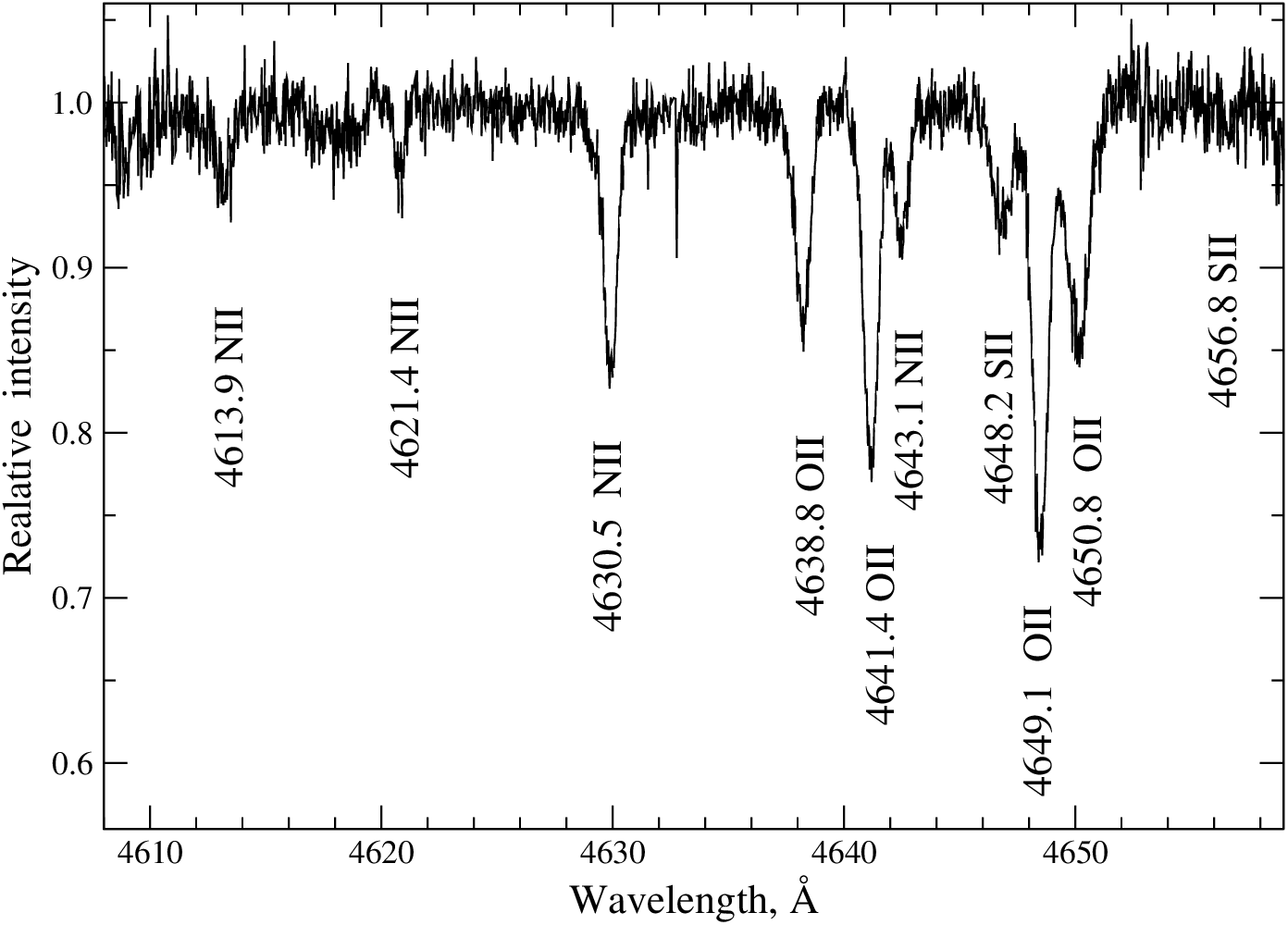}
\includegraphics[angle=0,width=0.45\textwidth,bb=0 0 750 600,clip]{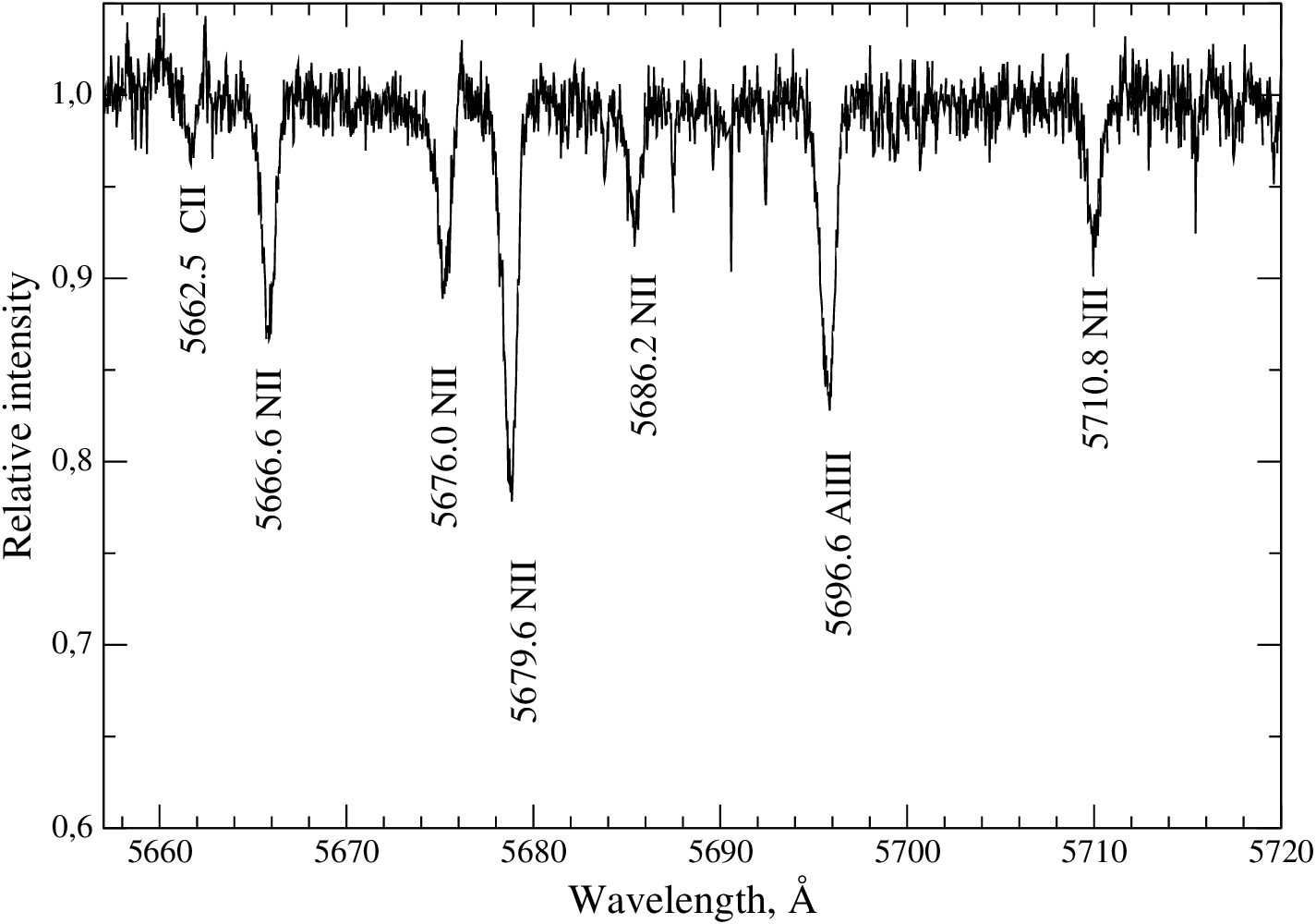}
}
\caption{Fragments of the spectrum of IRAS\,01005 taken on 29 May 2013 with photospheric type
   absorptions of the CNO triad,  S{\sc ii}, and Al {\sc iii}.}
\label{Frag593}
\end{figure}

\begin{figure}
\includegraphics[angle=0,width=0.45\textwidth,bb=0 0 700 700,clip]{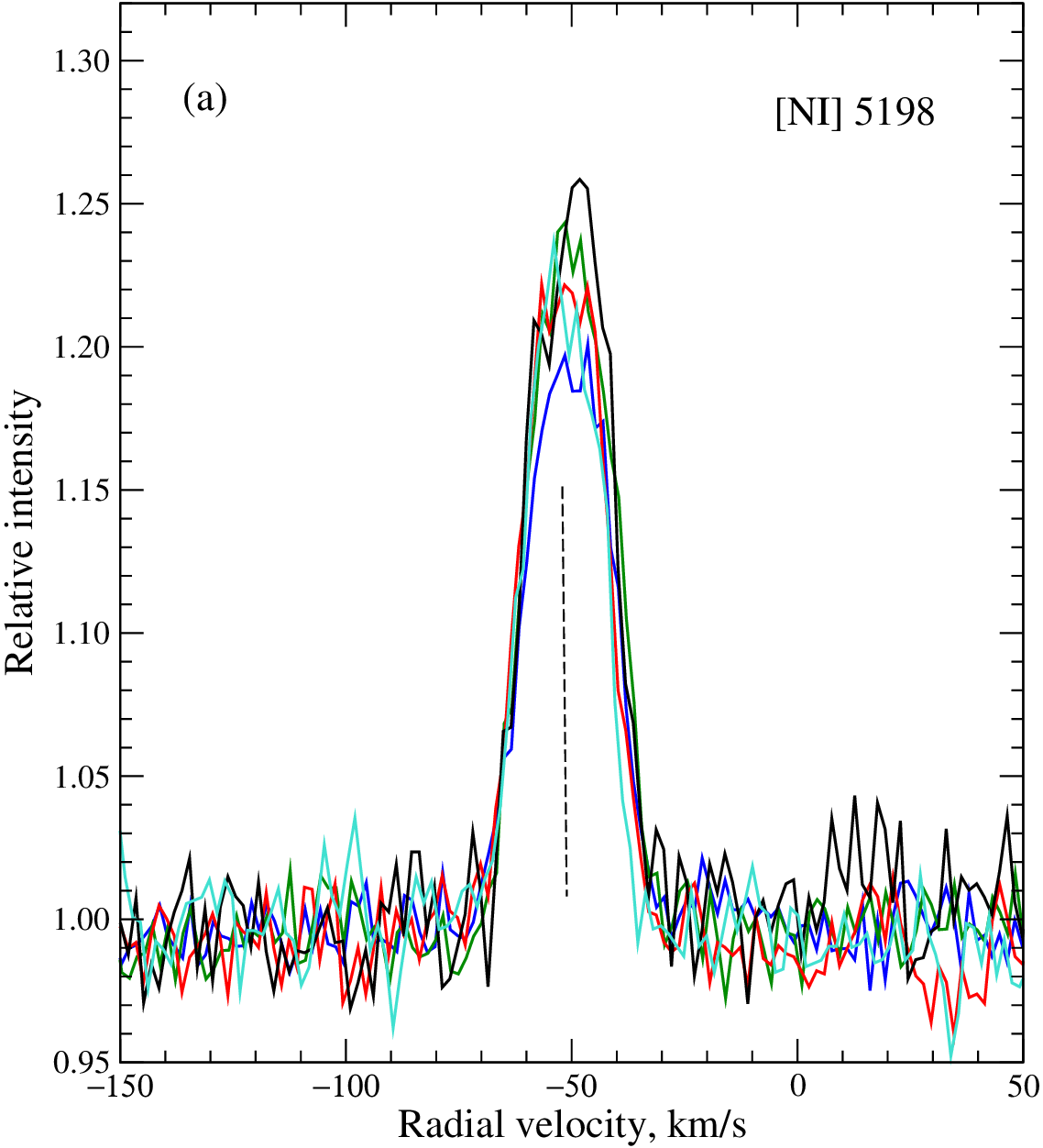}
\includegraphics[angle=0,width=0.45\textwidth,bb=0 0 700 700,clip]{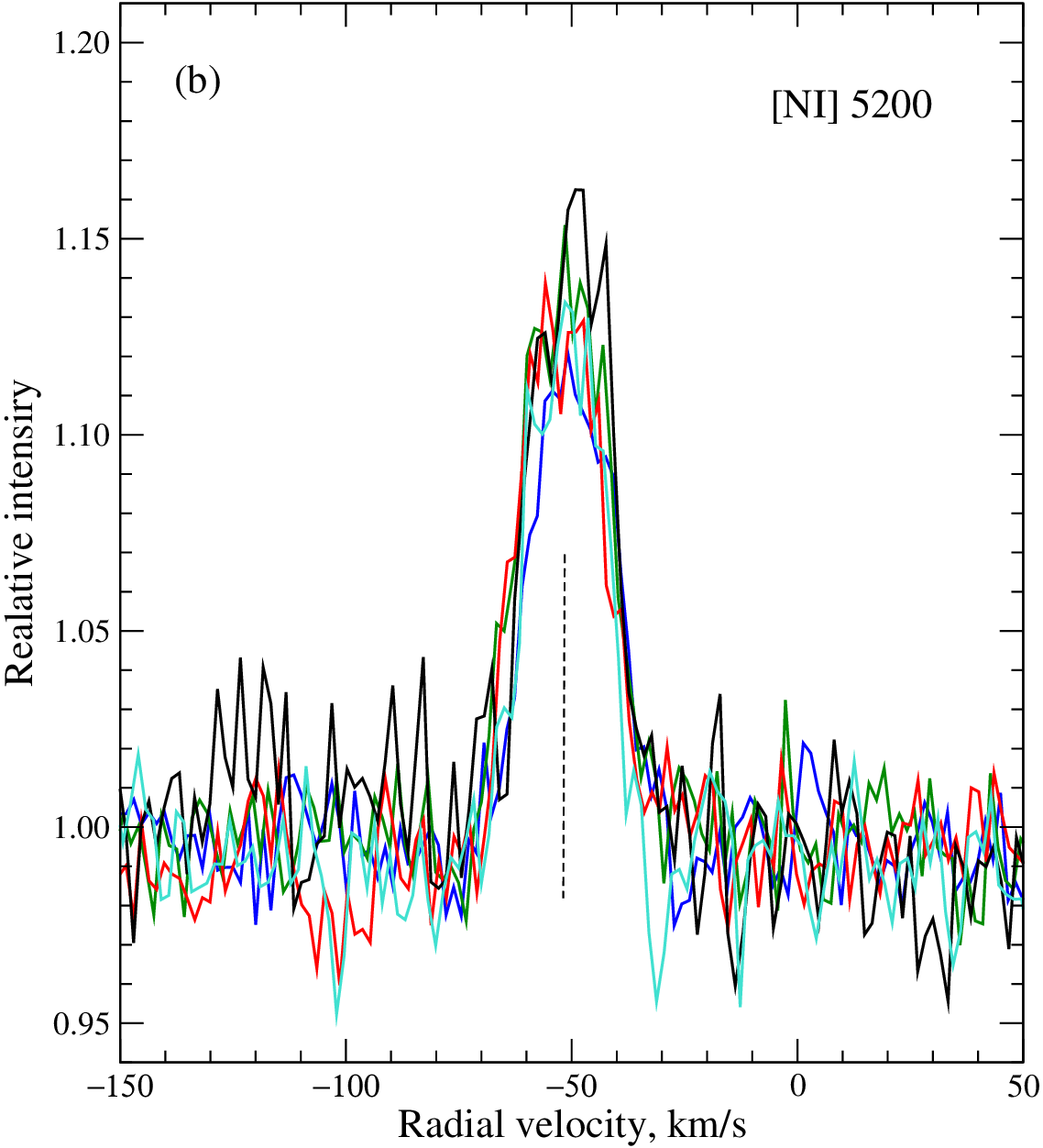}
\includegraphics[angle=0,width=0.45\textwidth,bb=0 0 700 700,clip]{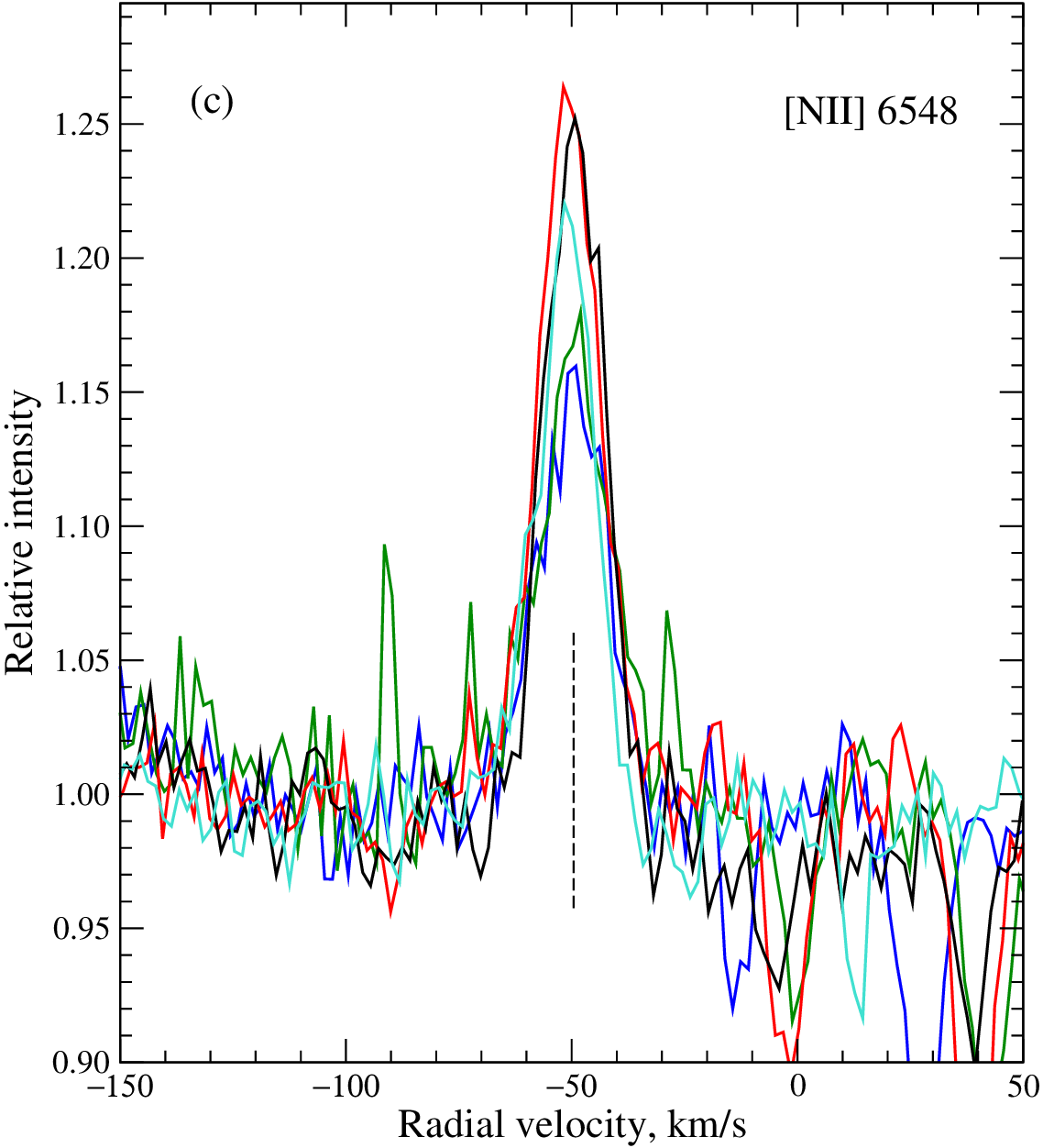}
\includegraphics[angle=0,width=0.45\textwidth,bb=0 0 700 700,clip]{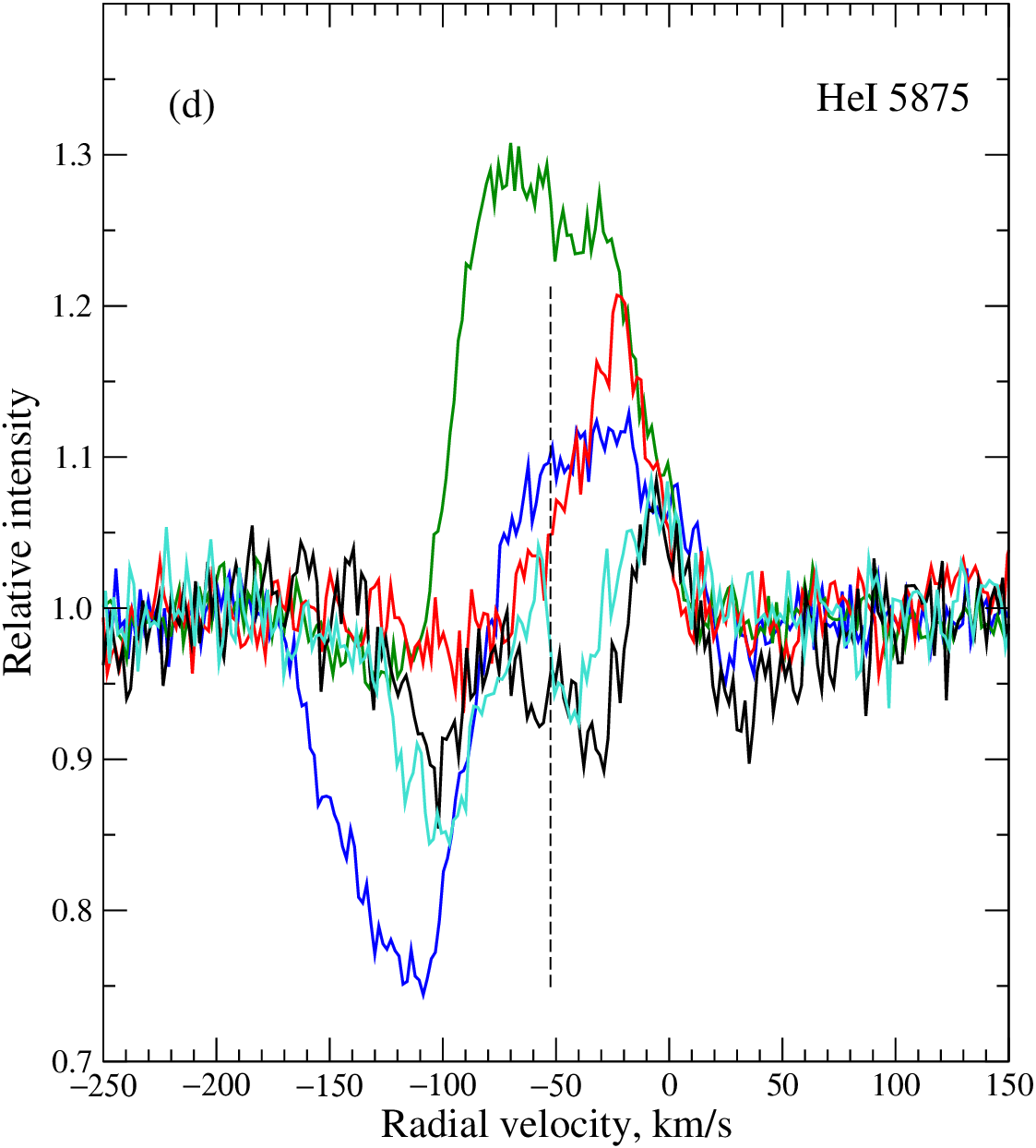}
\caption{Panels (\textbf{a}--\textbf{c}) show the forbidden lines of [N\,{\sc i}] 5198 and 5200\,\AA\, and
[N\,{\sc ii}]\,6548 \AA\, in the spectra taken on 29~May~2013 (blue), 21~August~2013 (green), 11~September~2025 (red),
4~October~2025 (black), and 27~January~2026 (turquoise), respectively.
Panel (\textbf{d}) shows the He\,{\sc i}~5876\,\AA\, line profiles for the same dates. The vertical dashed line  corresponds to the systemic velocity V$_{\rm sys}=-51.4$\,km\,s$^{-1}$. Telluric lines were not removed from the
plots.}
\label{Emis-forb}
\end{figure}

Detailed line identification in a broad wavelength range of the spectrum of IRAS\,01005 was published by~\citet{2014AstBu..69..439K}.  Symmetric photospheric absorptions (e.g., CNO triad ions, numerous weak
lines of Si\,{\sc ii}, Si\,{\sc iii}, S\,{\sc ii}, S\,{\sc iii}  and the iron group metals) coexist
with narrow forbidden emission lines (mainly [Fe\,{\sc ii}], [N\,{\sc i}], [N\,{\sc ii}], [O\,{\sc i}],
and [S\,{\sc ii}]) that form in optically thin layers of the envelope. Examples of typical fragments
of the spectrum  with photospheric  absorption lines are shown in Figure~\ref{Frag593}, while examples
of forbidden line profiles are shown in panels (a), (b), and (c) of Figure~\ref{Emis-forb}.

The intensity of narrow forbidden lines (with full-widths at half maximum of $\le$$10$\,km\,s$^{-1}$)
in the spectrum of IRAS\,01005 is not strong. It typically does not exceed 10\% above the local continuum
level, but these lines provide good accuracy of the radial velocity measurements. As follows from the data
shown in Table\,\ref{tab1}, the standard error for every observing night does not exceed 0.15\,km\,s$^{-1}$,
and the average V$_{\odot}$(emis-Forb)$=-51.36$\,km\,s$^{-1}$ with a semi-amplitude of $0.26$ km\,s$^{-1}$
for the five observing dates. Panels (a), (b), and (c) of Figure~\ref{Emis-forb}, where the profiles of the
strongest forbidden emissions  are shown, provide a good illustration of their stationary positions and
the weak temporal variability of their peaks. A~comparison of the data taken in 2025$\div$2026 with the earlier measurements~\citep{2002A&A...392..143K,2014AstBu..69..439K} confirms the conclusion  that the positions of these emissions, which form in an optically thin region of the circumstellar envelope, do not vary.

Panel (d) of Figure~\ref{Emis-forb}  shows a complex profile of the He\,{\sc i}~5875\,\AA\, line, which forms
in a transitional  region between the stellar atmosphere and the envelope in a wide range of radial velocities
and demonstrates a special behavior.    Significant variability in the line profile is due to the inhomogeneity
and asymmetry of the envelope and the instability of the wind.  Such behavior of this line profile is a good
illustration of a family of  absorption--emission spectral features. A low intensity of  the latter as well
as the transformation of their profiles by overlapping absorption components make it difficult to measure their
positions. As a result, the positional uncertainties shown in the last column of Table\,\ref{tab1} are much
larger compared to those of the radial velocity measurements from the positions of the forbidden lines and
photospheric absorptions.

The first column of Table\,\ref{tab1} shows the radial velocities of the photospheric absorption lines measured
with high precision. We stress that the majority of the pure absorptions (with no additional features) were
identified as features of ions, such as N\,{\sc ii}, O\,{\sc ii}, S\,{\sc ii}, and Al\,{\sc iii}.
Table\,2 in~\citet{2014AstBu..69..439K} presents the depths and radial velocities of most of these absorptions.
The new observations taken in 2025$\div$2026 revealed additional and more significant evidence of radial velocity
variability in a range of V$_{\odot}$ from $-20$ to $-68$\,km\,s$^{-1}$. Therefore, in the five reported
observing nights, we detected variability in the radial velocity of the absorption lines of
V$_{\odot}$(aver-abs)$=-40.36$\,km\,s$^{-1}$ with a semi-amplitude of $\sim$8\,km\,s$^{-1}$. Note that
all the radial velocities of the photospheric absorptions observed in 2000$\div$2013~\citep{2002A&A...392..143K,2014AstBu..69..439K} fall in the  same range between $-20$ to $-68$\,km\,s$^{-1}$,
and V$_{\odot}$(abs) in the spectrum taken on 18 January 2005 are very close to those
measured in the spectrum taken on 27 January 2026. At the same time, the radial velocity of
the forbidden lines, V$_{\odot}$(aver-Forb)$=-51.36\pm0.26$\,km\,s$^{-1}$, remains stable.

A significant increase in the radial velocity of photospheric absorptions of 20\,km\,s$^{-1}$ within
3 weeks,  from 11~September~2025 to 4~October~2025, is notable. Our previous data taken in 2000$\div$2013~\citep{2002A&A...392..143K,2014AstBu..69..439K}   show the same variability range for these lines,  while
the average radial velocity of the forbidden lines remains stable, V(aver-Forb)$=-51.36\pm0.26$\,km\,s$^{-1}$.
Multiple high-resolution spectroscopic observations are needed within a few months to refine the amplitude,
measure  the period, and better understand the reasons for the variations.

We consider the detection of the non-stationary character of the wind components of the H$\alpha$ and H$\beta$
lines to be a new result.        A dramatic change in these line profiles occurred in the spectra taken in
2025$\div$026. In particular, the wind component was absent on 11~September~2025 but observed again on
4~October~2025. We also note an unusual profile of the H$\beta$ line on 29~May~2013 (the blue line in the
right panel of Figure~\ref{Halpha5}) with a complex shape of the wind component.

Overall, judging by the features of the optical spectrum, the hot central star of IRAS\,01005 with a compact
dusty envelope is undoubtedly       a member of the B[e] object community. However, there is no certain
answer to the question of its evolutionary status. A high T$_{\rm eff}$ and the presence of forbidden Fe\,{\sc ii},
[Fe\,{\sc ii}] and [O\,{\sc i}] emission lines, as well as strong Balmer emission lines with P\,Cyg-type profiles,
allow us to suggest a similarity between IRAS\,01005 and a hot star [MA93]\,1116 in the Small Magellanic Cloud retained.
The results of low-resolution optical spectroscopy were published by~\citet{2007ApJ...670.1331W}, who determined a
T$_{\rm eff}$ of  19\,000~K and a luminosity of $\log(L/L_{\odot})\approx$4.4, and suggested a status of a
B[e] supergiant for this object.

Another object with similar properties of the optical spectrum to those of IRAS\,01005 is LS{\sc iii}\,+52$^{\circ}$24,
a hot star  located at a Galactic latitude of b=$-1\degr96$  and an optical counterpart of the IR source
IRAS\,22023+5249. The main features of  the spectrum of the latter object (very strong intensity of hydrogen
emission lines, He\,{\sc i} lines with variable profiles, signs  of atmospheric pulsations, and the presence of
forbidden lines of light metals) allowed~\citet{2022ARep...66..429K} to suspect that this star belonged to
B[e] supergiants. However, the entire set of features of LS{\sc iii}\,+52$^{\circ}$24, which include a moderate
luminosity, peculiarities of the chemical  composition according to~\citet{2012MNRAS.421..679S} and a  large
radial velocity, more likely correspond to a hot post-AGB star status. The spectrum of LS{\sc iii}\,+52$^{\circ}$24
may be an example of the phenomenon of  supergiant mimicry (see~\citep{2018ARep...62...19K} for a detailed
description).

A close analog of IRAS\,01005 is AS\,314, a star of a later spectral type (A0). A set of photometric and spectral
features of the latter  (an excess of far-IR radiation, P\,Cyg-type profiles of the H$\alpha$ and H$\beta$ lines,
forbidden lines in the optical spectra) revealed  over the course of multi-year observations
led~\citet{2025Galax..13...17B} to no unambiguous choice between the following possibilities: a massive evolved
star (a hypergiant or a Luminous Blue Variable) or a low-mass star in transition from the AGB to the Planetary
Nebula stage. These authors tend to accept the massive star status for AS\,314.

Similar doubts on the status of the object were raised and a similar preferred conclusion about IRAS\,01005
was made at earlier stages of its investigation~\citep{2002A&A...392..143K}. With the new spectra and
additional conclusions reported above, we can classify the object as a supergiant based on the chemical
composition of the atmosphere, which is inconsistent with those of post-AGB stars, taking into account a fast
and unstable wind and large widths of the photospheric lines. An important reference to a large initial mass
of IRAS\,01005 can be abundant features of nitrogen and its ions. The presence of forbidden lines of [N\,{\sc i}]
and [N\,{\sc ii}], which form in the envelope, may point to the synthesis of nitrogen in the core of a massive
star in the past and a subsequent transfer into the atmosphere and the circumstellar medium. However, a
luminosity of $\log(L/L_{\odot})$=3.6 for IRAS\,01005, derived using the Gaia~DR3 parallax~\citep{2021A&A...649A...1G}
and corrected for the interstellar extinction calculated from the diffuse interstellar band strengths~\citep{2002A&A...392..143K}, is much lower than the boundary suggested for B[e] supergiants
in the review by~\citet{2019Galax...7...83K}. We note that the star in IRAS\,01005 with a luminosity of
$\log$ L/L$_{\odot} = 3.39$ was included in a list of Galactic post-AGB stars~\citep{2022MNRAS.516L..61O} based
on the Gaia DR3 parallax and  assigned category ``I''---most likely a post-AGB object---by~\citet{2015MNRAS.447.1673V}.

The parameters of IRAS\,01005, including the presence of forbidden features of [N\,{\sc i}] and [N\,{\sc ii}],
allow us to suggest  the post-AGB status with a high initial mass, while the Hot-Bottom Burning (HBB) process
can provide excess nitrogen~\citep{1993ApJ...416..762B}. A star undergoing a short-term HBB stage possesses a
unique internal structure, which includes a degenerate CO-core surrounded by two alternatively active layers of
helium located above the inert core and of hydrogen located below the convective envelope. The intershell process
of helium burning is unstable, accompanied by flares and mixing episodes, and leads to a chain of neutronization
reactions of metallic nuclei (the slow s-process) and transfer of the newly created material to the convective
envelope and  the stellar surface.  A detailed description of these episodes, commonly called the third
dredge-up (TDU), is presented  in~\mbox{\citet{2005ARA&A..43..435H,2022Univ....8..170B,2007PASA...24..103K}}.
We note that a sample of similar post-AGB stars with complex envelopes and atmospheric chemical composition
altered in the course of the s-process and further mixing was studied using high-resolution spectroscopy
carried out at the BTA telescope (see examples in~\citep{1995MNRAS.272..710K,1999A&A...345..905K,2013AstL...39..765K,2015AstL...41...14K}).

Quiet hydrogen burning in a layer is established after the end of TDU. At the same time, as stressed by~\citet{2007MNRAS.378.1089M}, nitrogen synthesis occurs in the entire volume of the upper hydrogen
envelope of a post-AGB star, which includes its outermost layers, where nitrogen synthesis occurs by
transformation of C$^{12}$ atoms into N$^{14}$. The results of this evolutionary phase suggest the efficiency
of HBB  in stars with initial masses of 4$\div$5\,M$_{\odot}$~\citep{2015MNRAS.450.3181V}. The consequences
of the HBB process at such a late stage of the star's evolution can be significant changes in its surface
chemical composition. In particular, a C-rich post-AGB star can further exist as  an N-rich one. Therefore,
investigation of the spectrum of IRAS\,01005 gives us additional confirmation of the old conclusion by~\citet{1998A&A...340..117L} about the creation of the B[e] phenomenon in objects at different
evolutionary stages but with close physical conditions in the circumstellar medium.

\section{Conclusions}\label{sec4}

We detected variations in the wind component of the H$\alpha$ and H$\beta$ emission lines by expanding
spectroscopic monitoring of  IRAS\,01005.  The positional stability of a sample of forbidden emissions
of the CNO triad, Si and species of the iron group led  to an accurate determination of the systemic
radial velocity of V$_{\odot}$(sys)=$-51.36\pm0.26$ km\,s$^{-1}$.

The high accuracy of the positions of spectral features enabled the detection of the radial velocity
variations of the photospheric type absorptions with respect to an average value of
V$_{\odot}$(aver-abs)=$-40.36$\,km\,s$^{-1}$ with a preliminary estimate of the  semi-amplitude of
the variations of 8.0\,km\,s$^{-1}$.    The conclusion of the presence of pulsations in the atmosphere
of the central star or the existence of a stellar companion has been reached. Continuation of
high-resolution spectroscopic monitoring is needed to constrain the amplitude and period of the
variability, as well as the reason for it.

A reliable measurement of the star's luminosity and distance is the main task for constraining
the evolutionary status of a star with  a complex envelope.  The observed set of properties of
IRAS\,01005 suggests a status of a post-AGB star that has undergone the HBB phase.

\vspace{6pt}
\section*{FUNDING}
This research was completed within the framework of the state request to the Special Astrophysical
Observatory of the Russian Academy of Sciences from the Ministry of science and higher education of
the Russian Federation.
The project was partly carried out within the framework of  Project No. BR24992759
''Development of the concept for the first Kazakhstani orbital cislunar telescope---Phase I'',
financed by the Ministry of Science and Higher Education of the Republic of~Kazakhstan).

\vspace{6pt}
\section*{acknowledgments}This research has made use of the SIMBAD database, operated at
CDS, Strasbourg, France;  the Astrophysics Data System, funded by NASA under Cooperative
Agreement 80NSSC21M0056.;  VALD databse of atomic data; and  data from the European Space
Agency (ESA) mission Gaia.

\end{document}